\documentclass[12pt,preprint]{aastex}
\begin{document}
\title{\Large{Spitzer IRAC Images and Sample Spectra of Cassiopeia A's Explosion}}
\title{\large{ }}
\title{\large{ }}
\pagestyle{plain}
\textheight8.95in

\author{{Jessica A. Ennis, Lawrence Rudnick}\altaffilmark{1}}
\email{jennis@astro.umn.edu,larry@astro.umn.edu}

\author{{William T. Reach}\altaffilmark{2}}
\email{reach@ipac.caltech.edu}

\author{{J. D.  Smith}\altaffilmark{3}}
\email{jdsmith@as.arizona.edu}

\author{{Jeonghee Rho}\altaffilmark{2}}
\email{rho@ipac.caltech.edu}

\author{{Tracey DeLaney}\altaffilmark{4}}
\email{tdelaney@cfa.harvard.edu}

\author{{Haley Gomez}\altaffilmark{5}}
\email{haley.gomez@astro.cf.ac.uk}
\author{{Takashi Kozasa}\altaffilmark{6}}
\email{kozasa@ep.sci.hokudai.ac.jp}

\altaffiltext{1}{Astronomy Department, University of Minnesota, Minneapolis, MN 55455}
\altaffiltext{2}{Spitzer Science Center, California Institute of Technology, MS 220-6, Pasadena, CA 91125}
\altaffiltext{3}{Steward Observatory, 933 N. Cherry Ave, Tucson, AZ 85712}
\altaffiltext{4}{Harvard-Smithsonian Center for Astrophysics, 60 Garden St, MS-67, Cambridge, MA 02138}
\altaffiltext{5}{School of Physics and Astronomy, University of Wales, Cardiff, Wales, UK}
\altaffiltext{6}{Department of Cosmosciences, Graduate School of Science, Hokkaido University, Sapporo,  060-0810, Japan}

\begin{abstract}
We present \textit{Spitzer} IRAC images, along with representative 5.27 to 38.5~\micron~ IRS spectra of the Cassiopeia~A supernova remnant. We find that various IRAC channels are each most sensitive to a different spectral and physical component. Channel~1 (3.6~\micron) provides an excellent match to the radio synchrotron images.  Where Channel~1 is strong with respect to the other IRAC channels,  the longer-wavelength spectra show a broad continuum gently peaking around 26~\micron, with weak or no lines.   We suggest that this is due to un-enriched progenitor circumstellar dust behind the outer shock, heated and potentially processed  by the photons and electrons from the shock.  Where Channel~4 (8~\micron) is bright relative to the other IRAC channels, the long-wavelength spectra show a strong, 2-3~\micron-wide peak at 21~\micron, likely due to silicates and proto-silicates. Strong ionic lines of  [Ar II], [Ar III], [S IV] and [Ne II] also appear in these strong Channel 4 regions. We suggest that in these locations, the dust and ionic emission originate from the explosion's O-burning layers.  The regions where Channels 2 (4.5~\micron) and 3 (5.6~\micron) are strongest relative to Channel~4 show a spectrum that rises gradually to 21~\micron, and then flattens or rises more slowly to longer wavelengths, along with higher ratios of [Ne II] to [Ar II].  We suggest that the dust and ionic emission in these locations arise primarily from the C- and Ne- burning layers.  There are no bright lines in the Channel 3 spectra themselves, although strong line emission, perhaps from [Fe II], Br~$\alpha$ or CO, must contribute significantly to the Channel 2 brightness.

All of these findings are consistent with asymmetries deep in the explosion, producing variations in the velocity structure in different directions, but generally preserving the nucleosynthetic layering.  At each location, the dust and ionic lines in the mid-infrared, and the hotter and more highly ionized optical and X-ray emission are then dominated by the layer currently encountering the reverse shock in that direction.
\end{abstract}

\keywords{ supernova remnant: Cassiopeia A -- dust -- nucleosynthesis}
\clearpage

\section{Introduction}
Cassiopeia A (Cas A)  is the youngest supernova remnant (SNR) in our galaxy, thought to be the result of either a type Ib or IIn supernova explosion \citep{chevalier} occurring in 1671 \citep{thorstensen} at a distance of
3.4 kpc \citep{reed}. The remnant has been studied extensively
at many wavelengths, and is one of the brightest radio and X-ray sources in the sky.  Its primary structures are a 105\arcsec\ radius bright ring surrounded by a 150\arcsec\ radius low surface brightness plateau \citep{braun}. The outer plateau is bordered by a thin X-ray ring identified as the outer shock in the circumstellar medium (CSM), with the broader, brighter interior ring originating from stellar ejecta that have encountered the reverse shock \citep{gotthelf}.

The X-ray emission is characterized by a thermal spectrum containing emission lines from highly ionized atoms. Optical emission from the remnant is dominated by chemically-enriched knots. Infrared emission was previously known to contain thermal continua from heated dust, line emission from ionized atoms, and, shortward of about 5~\micron, synchrotron emission from electrons accelerated in shock regions \citep{jones,rho}. Submillimeter observations of Cas A prove difficult due to a molecular cloud complex along the line of sight and the presence of cold dust in the remnant is therefore still in question \citep{dunne,krause, wilson}. The radio emission from Cas A is synchrotron radiation \citep{ginzburg}.

Cas~A's structure and dynamics reflect different progress into the Sedov-Taylor evolutionary phase in different directions \citep{delrud}.  The bright ring remains illuminated as new, successively slower-moving ejecta encounter the reverse shock and are heated and ionized \citep{delaney04}.   A Doppler analysis of the X-ray gas, studies of the optical knots, and the large abundance ratio of $^{44}$Ti/$^{56}$Ni support an asymmetric explosion \citep{reed,hwanglaming,willingale,nagataki}. The  MIPS images \citep{hines} showed both the main X-ray jet \citep{hwangholtpetre} and a counterjet \citep{hwangmsec}, providing further evidence for explosion asymmetry.

The progenitor of Cas A is generally believed to have been a WN star (i.e., a Wolf-Rayet star with high nitrogen abundance), due to the high abundances of N and H in some of the Fast-Moving Knots \citep[FMKs,][]{kamper,fesenbecker}.  The hydrodynamical model of \citet{perezrendon} suggests a $29-30\rm ~M_{\odot}$ progenitor, while \citet{young} find the overall data are best fit by a $15-25\rm ~M_{\odot}$ progenitor that loses its hydrogen envelope in a binary interaction.  The pre-supernova wind produced a dense, clumpy medium \citep{chevalier} which is currently being shocked by the blast wave.  The highest density shocked clumps are seen in optical emission as slow-moving  Quasi-Stationary Flocculi \citep[QSFs,][]{vandenbergh}.

Rich in optical, X-ray and infrared emission lines from ionized atoms \citep{fesenHST,douv99,hwangholtpetre},  Cas A provides an important window into both quasi-equilibrium (pre-explosion) and explosive nucleosynthesis.  Each layer contains several different elements, with C and O produced in He-burning, Ne and Mg first appearing through C-burning, and O and Al added with Ne burning \citep{woosweav,woosjank}.  When O and Mg burn, the heavier elements, Si, S, Ar, Ca are then produced  and their burning products yield the Fe group elements.   In Cas~A, much of this layered nucleosynthetic structure has been preserved following the supernova explosion, e.g., with layers of nitrogen- sulfur- and oxygen- rich ejecta seen beyond the outer shock \citep{fesenN}.  There is also some evidence for mixing of the nucleosynthetic layers on various scales.  Optical ``mixed emission knots'' show both N and S lines, suggesting that high speed clumps of S ejecta penetrated through the outer N-rich layers \citep{fesenN}.   At infrared wavelengths, ISOCAM observations showed the presence of Ar and S in strong Ne knots, although the Ne and silicate emissions appeared anti-correlated \citep{douv99}. The X-ray line data show some large-scale overturning of ejecta layers \citep{hughes, hwanglaming}; iron in the SE region is further out and is moving faster than the Si/O regions.

 The variations in ionic species at different locations, as probed by optical, infrared, and X-ray observations, are sensitive to temperature, density and ionization state.  Thus, multiple wavelength detections of a species from different ionization states can help separate these effects from actual abundance variations \citep{vink}. This  provides one important motivation for our \textit{Spitzer} studies.  Density information can be derived from comparing different line strengths from the same ionization state, e.g. for [S III] \citep{houck}.  All of this information aids in reconstructing a picture of the inhomogeneities in the explosion, produced, e.g., through instabilities deep in the core \citep{foglizzo,blondin}.

Another  key motivation for the \textit{Spitzer} observations presented here  was to understand the production and destruction of dust in Cas~A.   One possible location for dust production is in optically thick clumps of ejecta, such as proposed by \cite{lucy} for SN1987A.  They suggest that most of the dust is contained in such optically thick knots, with small grains being distributed diffusely between them. Ejecta knots that are dense enough should remain largely intact with the passage of the reverse shock and blast-wave. If the dust contained in dense clumps is not in equilibrium with the more diffuse X-ray gas, then it may remain at a colder temperature.

 In Cas~A, \citet{dwek87} detected a strong excess of emission in \textit{Infrared Astronomy Satellite} (IRAS) observations at 12~\micron~ to 100~\micron, largely from dust swept up by the supernova blast wave (the outer shock). \textit{Spitzer} MIPS observations of Cas A \citep{hines} also detected this thermal dust emission from shocked circumstellar material. In addition, emission has been seen from a hot dust component associated with both the optical  \citep{dwek87,fesenHST} and the X-ray ejecta \citep{dou}.  Dust continua found in Cas~A with the \textit{Infrared Space Observatory} (ISO) were fit at 21~\micron\ with proto-silicates \citep{arendtiso} or  MgSiO$_{3}$ and SiO$_{2}$ \citep{dou}. Those data suggest that the dust is heated continuously, presumably by the hot X-ray emitting gas \citep{dwek87,hines}.  As noted above, the amount of cold dust associated with Cas~A is still uncertain.

In this paper, we report the results of {\it Spitzer Space Telescope} images made using the Infrared Array Camera \citep[IRAC,][]{irac} with brief supporting data from the  Infrared Spectrograph \citep[IRS,][]{jrhouck} and images from  other wavelengths.  We show evidence for different nucleosynthetic layers currently encountering the reverse shock in different directions. In subsequent papers, we will address the physical conditions  and dynamics of the gaseous material, and the detailed composition, temperature structure and mass estimates of the dust components.
% a fuller treatment of the IRS results \citep{Smith06}, the dust modeling \citep{rho06}, including the critical issues of condensation efficiency \citep{Mai06} and multiple temperature components \citep{dunne,morgan,krause}, the dynamics using Doppler shifts from multiple species \citep{Delaney06} and line emission from the remnant's interior \citep{Rudnick06}.

\section{Observations}

\begin{figure}
\begin{center}
%\plotone{IRAC1234a.eps}
\plotone{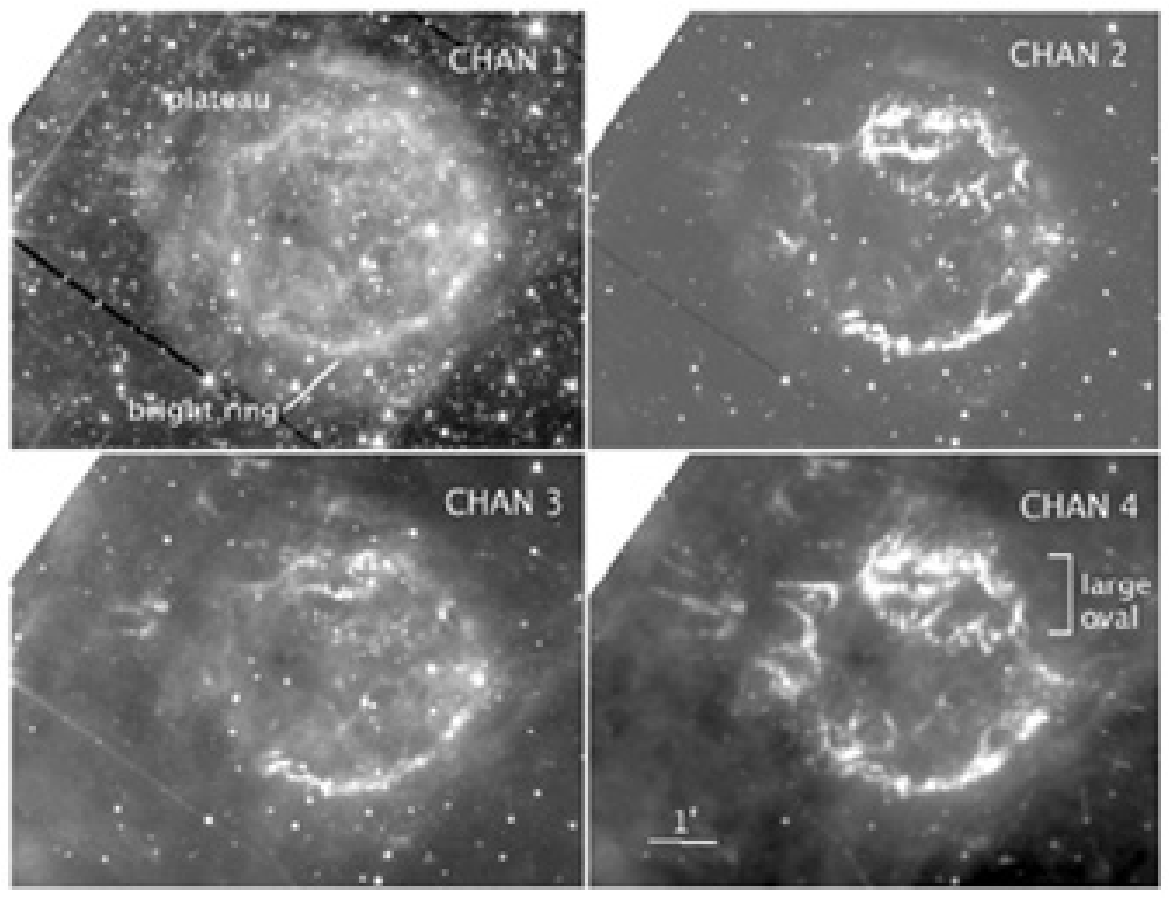}
\end{center}
\caption{IRAC images of Cassiopeia A. Channel 1: 3.2 -- 3.9~\micron. Channel 2: 4.0 -- 5.0~\micron.  Channel 3: 5 -- 6.4~\micron. Channel 4: 6.5 - 9~\micron. The bright ring marks the location of the large-scale reverse shock, and the outer edge of the
faint, diffuse plateau is the location of the forward shock. For full-resolution images, please see http://webusers.astro.umn.edu/\~{}jennis/iracpaper.html.}
\label{iracs}
\end{figure}

IRAC observations covered the entire Cas A supernova remnant, including the outer shock, jet and counterjet regions.
The IRAC images utilize four wide filters with central wavelengths of 3.6, 4.5, 5.6 and 8~\micron~ for Channels 1, 2, 3 and 4, respectively.  The data were taken on January 18, 2005.
%Channel 1 covers 3.19--3.94\,$\mu$m,
%Channel 2 covers 4.02--5.03\,$\mu$m,
%Channel 3 covers 5.02--6.44\,$\mu$m, and
%Channel 4 covers 6.45--9.38\,$\mu$m.
The observing strategy combined a mapping grid and dithers to yield
a depth of coverage of at least 18 pointings over the entire remnant,
with higher coverage in some overlap regions. At each pointing, a
0.6 and a 12 second frame was taken. The IRAC images have an angular
resolution of 2-2.5$''$ in Channels 1--2 and $\approx$~3$"$ in Channels 3--4. The data were processed
with the S11 version of the IRAC pipeline \citep{lowrance}.   The four IRAC images are shown in Figure \ref{iracs}.

The IRS Spectrograph  was used on January 13, 2005 to spectrally map the full remnant (with portions of the outer structures missing
from some slits), covering 5--15\,$\mu$m\ (Short-Low module, SL) and 15-38\,$\mu$m\ (Long--Low module, LL). Each included the two orders of wavelength, and the two long, low-resolution slits provided resolving powers of  64--128. The long-wavelength (15--38\,$\mu$m) spectra were taken in a single large map with $4\times 91$ pointings, using a single 6 second ramp at each position.  To achieve the spatial coverage  with the short-wavelength (5--15\,$\mu$m) slit, a set of four quadrant maps were made, two with $4\times 87$ pointings and two with $3\times 87$ pointings, using  a 6 second ramp at each position. The mapped area ranged from 6.26$'$ $\times$ 5.86$'$ (SL) to 11.0$'$ $\times$ 7.79$'$ (LL), with offsets between the maps produced in each of the two orders in each module of ~3.2$'$ (LL) and ~1.3$'$ (SL), along the slit direction. The effective overlap coverage of all modules and orders is ~4.9$'$ by ~5.8$'$.
%Further details of the spectral mapping observation will be presented in \cite{Smith06}.
The illustrative spectra presented here were processed with the S12 version of the IRS pipeline, using the Cubism package \citep{Smith06} to reconstruct the spectra at each position. They were extracted from the IRS cubes using areas from 10$\arcsec$ to 33$\arcsec$ across.  No detailed matching of spectral overlap amplitudes between the SL and LL detectors was done.  Occasional instrumental problems, such as spectral fringing at the short wavelength end of LL2, are not addressed  and do not affect the analysis presented here.

%The overall spectrum of the remnant is shown in Figure \ref{total}, along with the identification of various ionic lines, and the spectral regions where Channels 3 and 4 overlap with IRS coverage.
For comparisons with the IRAC images, we  also performed near-infrared observations with a narrow Pa~$\beta$ filter, using the Palomar 200-inch Wide-infrared Camera (WIRC).  The  Pa~$\beta$  filter is cented on 1.282$\mu$m with a  1\% width.  The data were taken  on Aug 15 and 16, 2005. The exposure time  was nine  times 90~seconds per sky position.

\section{Results}\label{results}
\subsection{IRAC images and comparisons to other bands}

The four IRAC images  in Figure \ref{iracs} each show the same overall structure of the remnant, including the bright ring, the
surrounding low surface brightness plateau, and the eastern jet.  The plateau region  and the internal filamentary emission are most apparent in Channel 1.   The large oval ring covering the northern third of the remnant is prominent in Channels 2 and 4, weak in Channel 3, and just visible in Channel 1.  Channels 3 and 4 show significant diffuse and patchy emission beyond the plateau, likely associated with the surrounding medium.

\begin{figure}
\begin{center}
%\plotone{color.eps}
\plotone{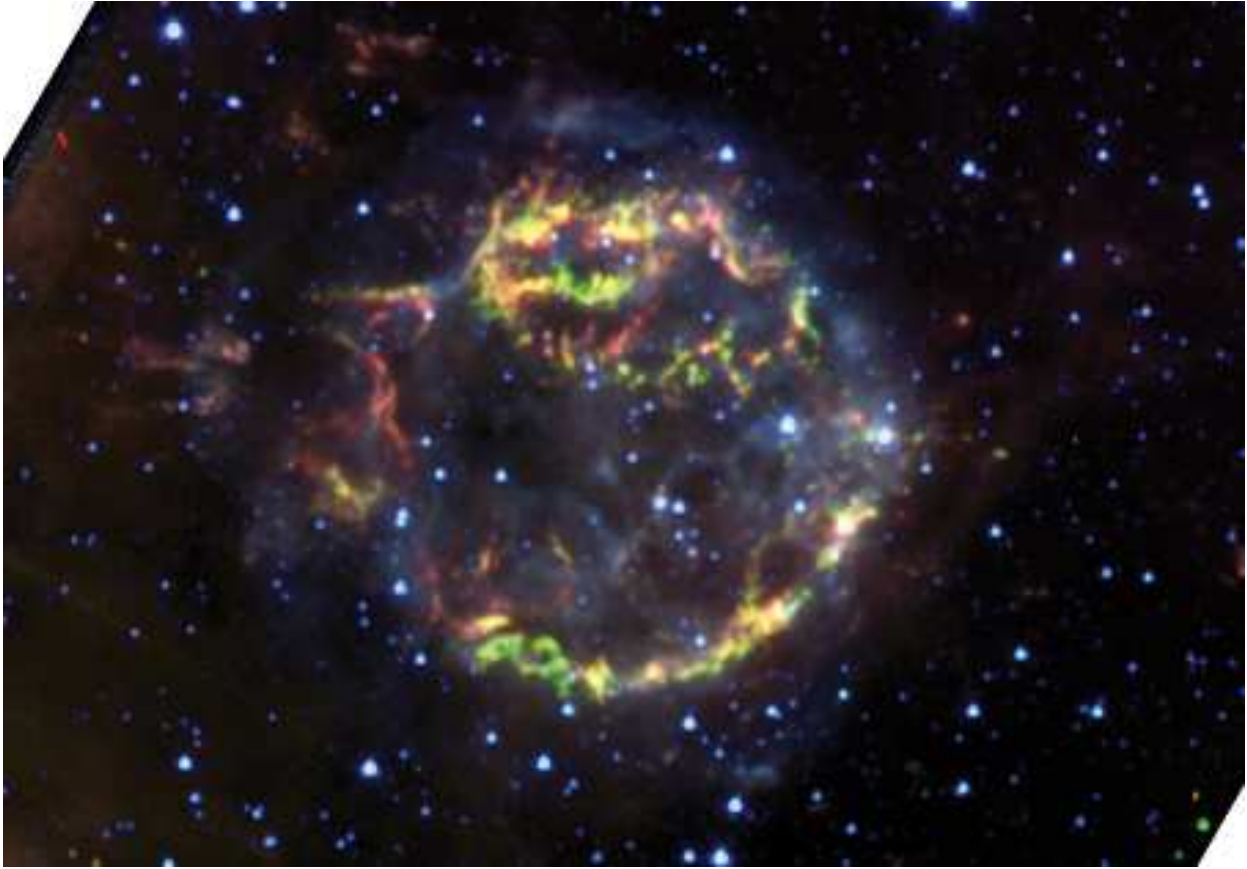}
\end{center}
\caption{Color combination of IRAC images.  Channel 1: blue;  Channel 2: green;  Channel 3: orange-yellow;  Channel 4: red. For full-resolution images, please see http://webusers.astro.umn.edu/\~{}jennis/iracpaper.html.}
\label{iraccolor}
\end{figure}

\begin{figure}
\begin{center}
%\plotone{alliracspecb.eps}
%\plotone{iracfour2.eps}
\plotone{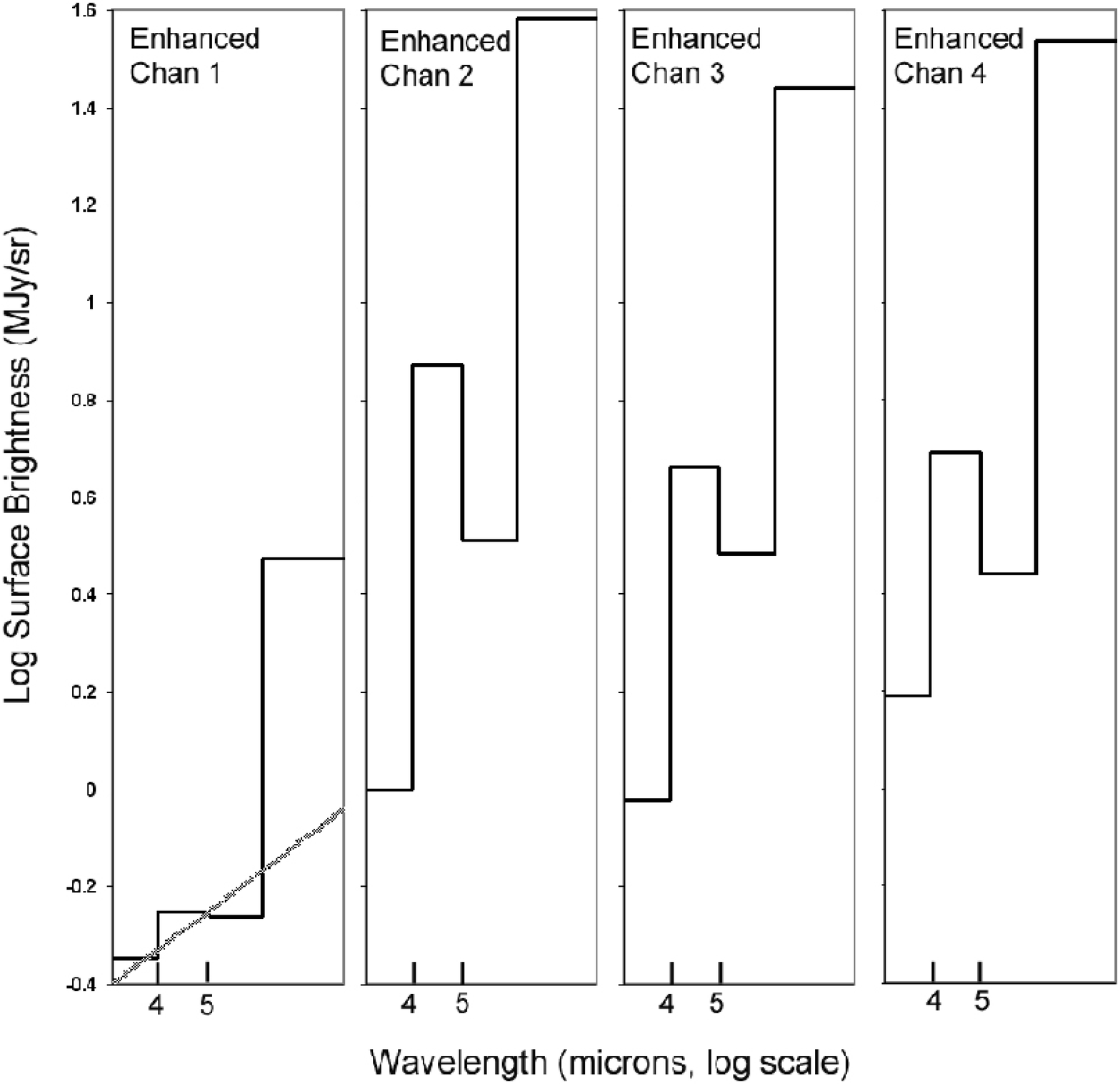}
\end{center}
\caption{Broadband IRAC spectra of locations in Cas A where each of the respective IRAC Channels are enhanced relative to the other channels.  An estimate of the synchrotron contribution is shown as a grey line superposed on the Channel 1 enhanced spectra.  For the other spectra, only Channel 1 can have a dominant synchrotron component because the brightness rises much too rapidly at longer wavelengths. For full-resolution images, please see http://webusers.astro.umn.edu/\~{}jennis/iracpaper.html.}
\label{iracspec}
\end{figure}

A color image combining all four  IRAC images  is shown in Figure \ref{iraccolor}. There are large spatial variations in the relative strengths of the IRAC channels, resulting in the broad range of IRAC ``colors".  In order to examine the colors more quantitatively, we isolated those regions where each channel was strongest with respect to Channel 4, and determined the mean surface brightness in those regions for each of the four channels.  The results are shown in Figure \ref{iracspec}. The very large jump to Channel 4 ($\approx$ 10) is likely caused by the presence of [Ar II] and [Ar III] emission in the IRAC band, as discussed further below.  Unfortunately, the IRS spectra do not cover Channels 1 and 2, and cover only part of Channel 3, so we cannot perform a quantitative analysis of these various IRAC colors; below, we suggest a few possible  contributors to Channels 2 and 3.

\begin{figure}
\begin{center}
%\plotone{synchrotron.eps}
\plotone{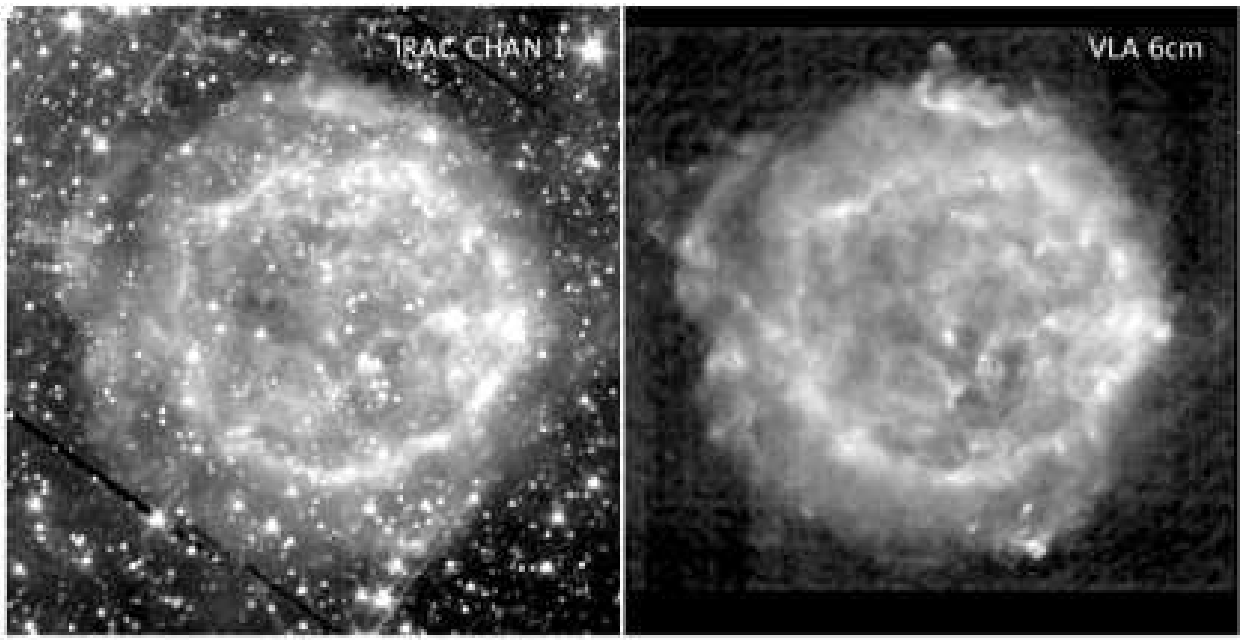}
\caption{Left: IRAC Channel 1;  Right: VLA radio image at 6cm from \cite{thesis}, 1.3'' resolution. For full-resolution images, please see http://webusers.astro.umn.edu/\~{}jennis/iracpaper.html.}
\label{synchrotron}
\end{center}
\end{figure}

In order to understand the origins of these IRAC color differences, we therefore first compare the images in each IRAC channel with those from other bands.  The IRAC Channel 1 image is very similar to that seen in the radio (Figure \ref{synchrotron}), with interior, ring, and plateau emission.   By contrast, the bright ring dominates the emission in  Channel 4,  mirroring Cas~A's appearance in the optical, in the 24~\micron~ MIPS image, and in X-ray line images (Figure \ref{ch4mips}). Channels 2 and 3 also are dominated by bright ring emission, although there are distinct differences in the brightness of various features between them and Channel 4.  Figure \ref{channel23} shows a comparison of Channels 2 and 3 with the [Fe~II] (18~\micron) and Pa~$\beta$ (1.3~\micron, with  some [Fe II] contamination) images. The results of these comparisons are all discussed in more detail below, after we have examined the correspondence  between the IRAC colors and the shapes of the IRS spectra.

We find no evidence for emission in the IRAC images from Cas~A's compact X-ray source \citep{tannan} against the variations in flux near the center of the remnant.  The 3-sigma upper limits are 50~$\mu$Jy for Channels 1 and 2, 100~$\mu$Jy for Channel 3, and 220~$\mu$Jy for Channel 4.

\subsection{IRS Spectra}

The 5.3 -- 38.5~\micron~ IRS spectrum for the full remnant is presented in Figure \ref{broadspec}, showing both the average continuum shape as well as Doppler-broadened ionic line emission. We indicate where the IRS spectral coverage overlaps IRAC Channels 3 and 4; there is no IRS coverage in the IRAC Channels 1 and 2 bands.  We also extracted spectra from 22 different regions in the remnant, chosen to explore a broad range of possible physical properties by comparing the IRAC color image to MIPS 24~\micron, X-ray, optical, and radio images.  These sample spectra showed a variety of relative line strengths and continuum shapes, especially in the relative strength of the peak around 21~\micron, as expected from the ISO work of \cite{dou}.  We found that the spectra fell into three major categories, as follows:  {\it Broad} -- showing a
 gentle peak around 10~\micron~ and rising to a very broad, gradual peak around 26~\micron, with little line emission; {\it Strong 21~\micron} -- showing a 2-3~\micron\ wide strong asymmetric peak at 21~\micron, similar to those studied with ISO \citep{dou}, along with strong lines of Ar, Ne, Si, S, and 26~\micron~ Doppler-blended Fe and O;  and  {\it Weak 21~\micron} -- rising gently through 21~\micron~and gradually becoming shallower to longer wavelengths, accompanied by stronger [Ne II], but relatively weaker [Ar II] lines.  Figure \ref{broadspec} shows the average shape for each of these classes; the variations of shape within each class are indicated by the grey bands.

We find a good correspondence between the IRS spectral shapes and the IRAC colors, as illustrated in Figure \ref{colorspec}.  When Channel 4 (with major contributions from [Ar II]) is strong with respect to Channel 2, the IRS spectra show the strong 21~\micron~ shape.  When Channel 4 is weaker, the ratio of Channels 2 and 3 distinguishes between broad spectra and weak 21~\micron~ spectra.  A relatively high ratio of Channel 1 to Channels 2 or 3 is also a good indicator of broad spectra.
Using the various IRAC colors, we can identify locations in the remnant where each IRAC channel in turn appears strongest compared to the other channels.  Table 1 summarizes the typical properties of regions where each respective channel is so enhanced.  We now look at each of the IRAC channels in turn, and discuss the likely origins of their emission.

\subsection{Strong IRAC Channel 1  (3.2 - 3.9~\micron)}

The Channel 1 image is shown in Figure~\ref{synchrotron}, along with a $\lambda$ 6cm
radio image from \cite{thesis}.  The detailed correspondence between these two images shows that Channel 1 is dominated by synchrotron emission. The synchrotron nature of the emission at 2.2~\micron~ was first suggested by \cite{gerardyfesen} based on its morphological similarity to the radio emission,  and then established both by its polarization \citep{jones} and brightness at levels expected from extrapolations of the radio spectrum \citep{rho,jones}. We now extend the detection of synchrotron radiation to the mid-infrared;  in IRAC Channel 1 we see the same  bright ring, faint plateau and filamentary structures as in the radio, at brightness levels comparable to those calculated from an extrapolation of the radio spectrum.  In the forward shock region, where IRAC Channel 1 is most enhanced relative to the other channels, synchrotron radiation makes substantial contributions to all the IRAC Channels (Figure \ref{iracspec}).  This region also shows significant 4-6 keV X-ray emission in the form of a thin rim of emission at the edge of the radio plateau.  The X-ray rim has been identified as marking the location of the outer (forward) shock \citep{gotthelf} and is likely dominated by synchrotron emission \citep{vink99, delaney04}.  A detailed analysis of the radio/infrared/X-ray spectral shape holds important clues to the relativistic particle acceleration mechanism \citep{ellison} but is beyond the scope of this paper.

Substantial emission from the forward shock region can also be seen in  the MIPS 24~\micron~ image \citep[Figure \ref{ch4mips},][]{hines}. However, the 24~\micron~ brightness reaches up to a factor of $\approx$50 above the values extrapolated from IRAC Channel 1, so it must be due to another emitting component.  IRS spectra of this spatial component show the characteristic ``broad" shape shown in Figure \ref{broadspec}.  Previous observations with ISOCAM  \citep{arendtiso} found these spectra rising to 18~\micron; with IRS, we now see a broad peak around 26~\micron, a smaller ``bump'' around 9~\micron, and little or no line emission.  The IRS spectrum of the forward shock region shown here can be approximately fit by a blackbody Planck function with a temperature of 113 K, multiplied by  the absorption efficiency calculated for grain models for R$_{\rm V}$~=~3.1 \citep{wein01}. The silicate emission feature between 9-11$\mu$m from the interstellar medium can also be seen; such silicate dust from the circumstellar/interstellar medium is expected in the forward shock region.

\subsection{Strong Channel 4  (6.5 - 9.4~\micron)}\label{ch4}

\begin{figure}
\begin{center}
%\plotone{OPTIRXb.eps}
\plotone{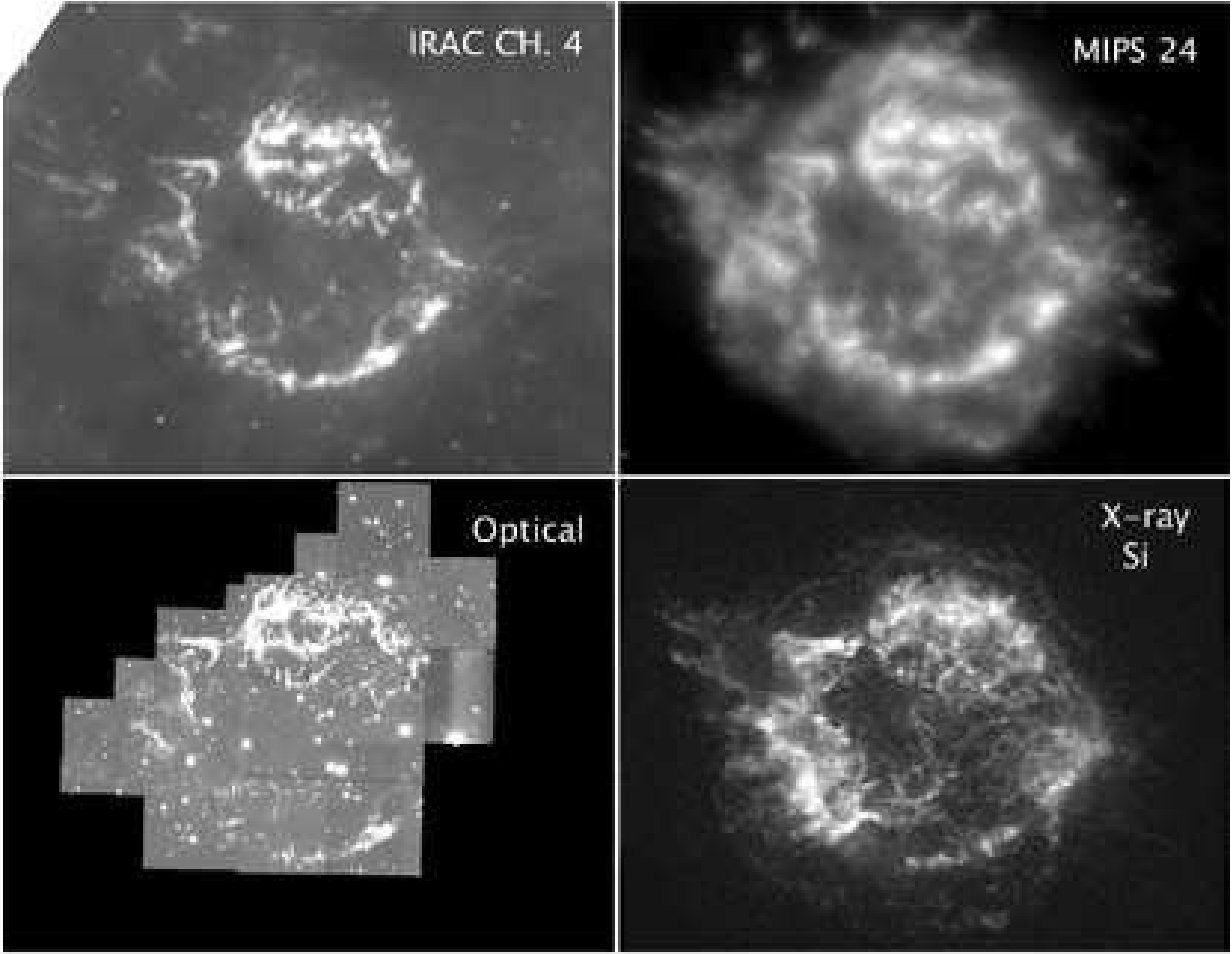}
\end{center}
\caption{Comparisons of IRAC Channel 4 with images from other wavelengths --  MIPS 24~\micron~\citep{hines}, HST WFPC2 F675W, \citep{fesenHST}, and Chandra ACIS X-rays for the Si line around 1.85 keV, using the epoch 2002 proper motion data set of \cite{delaney04}. For full-resolution images, please see http://webusers.astro.umn.edu/\~{}jennis/iracpaper.html.}
\label{ch4mips}
\end{figure}

The IRAC Channel 4 image (Figures \ref{iracs} and \ref{ch4mips} ) contains both continuum and line emission, with [Ar~II] at 6.99~\micron\ contributing approximately 40\% of the brightness, while [Ar~III] at 8.99~\micron~ contributes $\approx$3\%.  The 5--38~\micron~ IRS spectra of Channel 4 dominated regions also show  significant emission from  [Ne~II] (12.8~\micron) and [S~IV] (10.5~\micron). Strong lines from either [Fe~II] and/or [O~IV] (26~\micron), [Si~II] (34.8~\micron) and [S~III] (33.5~\micron), are also seen, but these are not exclusive to bright Channel 4 regions.  In most locations associated with the bright ring, Channels 2 and 3 are weak with respect to Channel 4 (e.g., as seen in Figure \ref{iracspec}).

The shape of the continuum of bright Channel 4 regions is characterized by a 2-3~\micron\ wide peak at 21~\micron, similar to that seen by \citet{arendtiso}  who  modeled it as due to magnesium proto-silicates.  This 21~\micron~ peak thus leads to an excellent correspondence between Channel 4 and the MIPS 24~\micron~image (Figure \ref{ch4mips}). However, the MIPS image also shows strong emission from the outer shock, whose spectra peak around 26~\micron, without any corresponding strong Channel 4 emission.

Many of the details of the Channel 4 emission can also be seen in the optical HST WFPC2 images; Figure \ref{ch4mips}  shows [S~II], [O~II] and [O~I] emission from dense shocked ejecta in the F675W image \citep{fesenHST}.  Some of the large-scale differences between the Channel 4 and WFPC2 images are due to variations in optical extinction.  The 0.3 - 10 keV X-ray emission \citep{hwangholtpetre} also shows some similarities to the Channel 4 image; a better correspondence is seen in the X-ray Si (shown in  Figure \ref{ch4mips})and S emission.  Ionic lines from lower ionization states of these two elements are seen in the IRS spectra of strong Channel 4 regions.

The ratio of [Ar~II] to [Ne~II] is highest in the bright Channel 4 regions, partly due to selection. In order to quantify this, we created continuum-subtracted line maps around the [Ne~II] (12.8~\micron) and [Ar~II] (6.99~\micron) images, and calculated the average ratio of those lines (see Table 1) in regions where the Channel 4 brightness was above 25 MJy/sr.  This will serve as a standard of comparison for the [Ar~II]/[Ne~II] ratio in places where the other IRAC channels are strongest relative to Channel 4.

\subsection{Relatively Strong IRAC Channels 2 (4 -- 5~\micron) and 3 (5 -- 6.4~\micron) }\label{ch3}

\begin{figure}[t]
%\plotone{IRAC23mos.eps}
\plotone{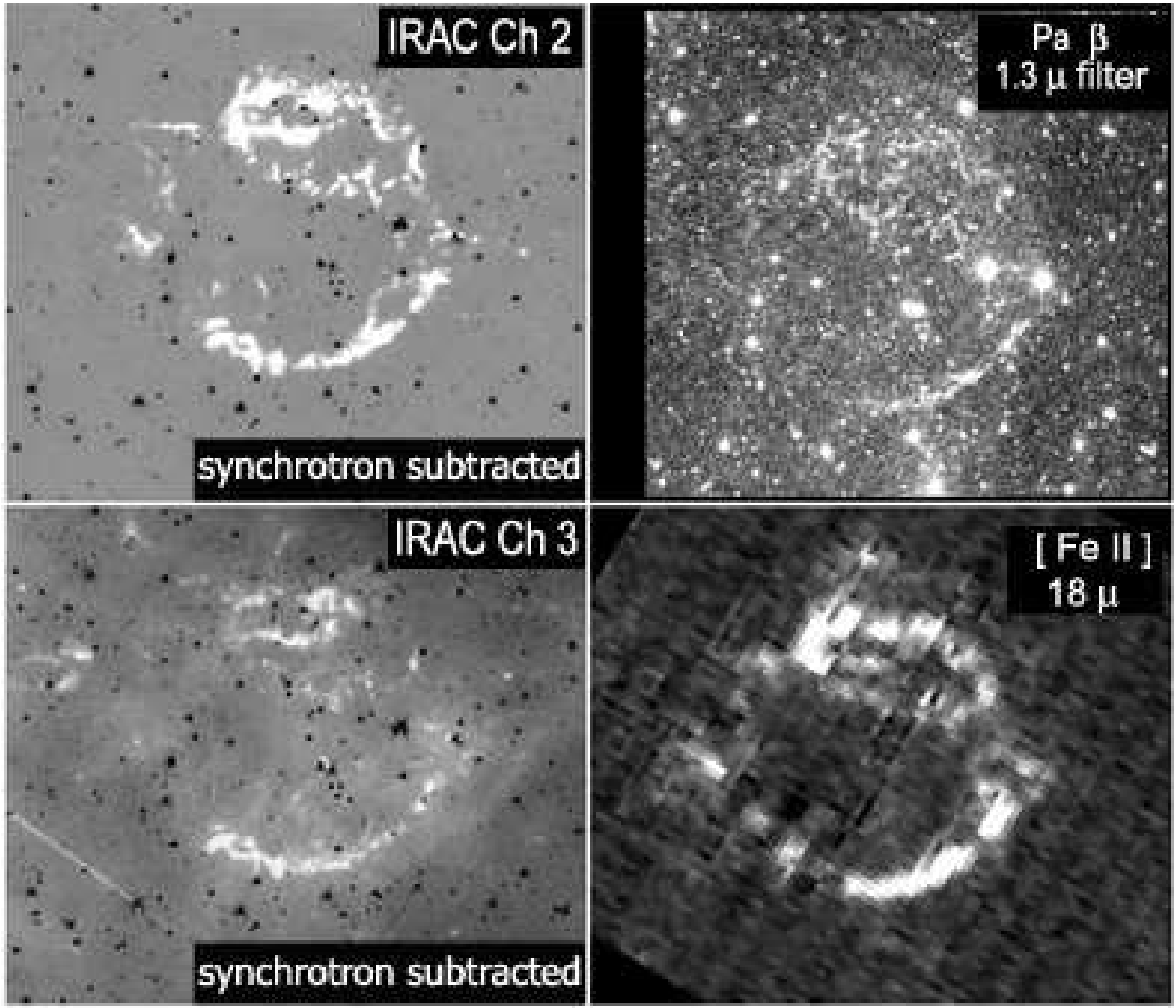}
\caption{Synchrotron subtracted images of IRAC Channels 2 (top) and 3 (bottom) for comparison to possible indicators of lines in 4~-~5~\micron~ range -- Pa~$\beta$, (with some [Fe II] contamination) at $\approx$1.3~\micron, and  [Fe II] image at $\approx$18~\micron. For full-resolution images, please see http://webusers.astro.umn.edu/\~{}jennis/iracpaper.html.}
\label{channel23}
\end{figure}

\begin{figure}
\begin{center}
\epsscale{1.0}
%\plottwo{newwhole.eps}{newbroad.eps}
%\plottwo{newstrong.eps}{newweak.eps}
\plottwo{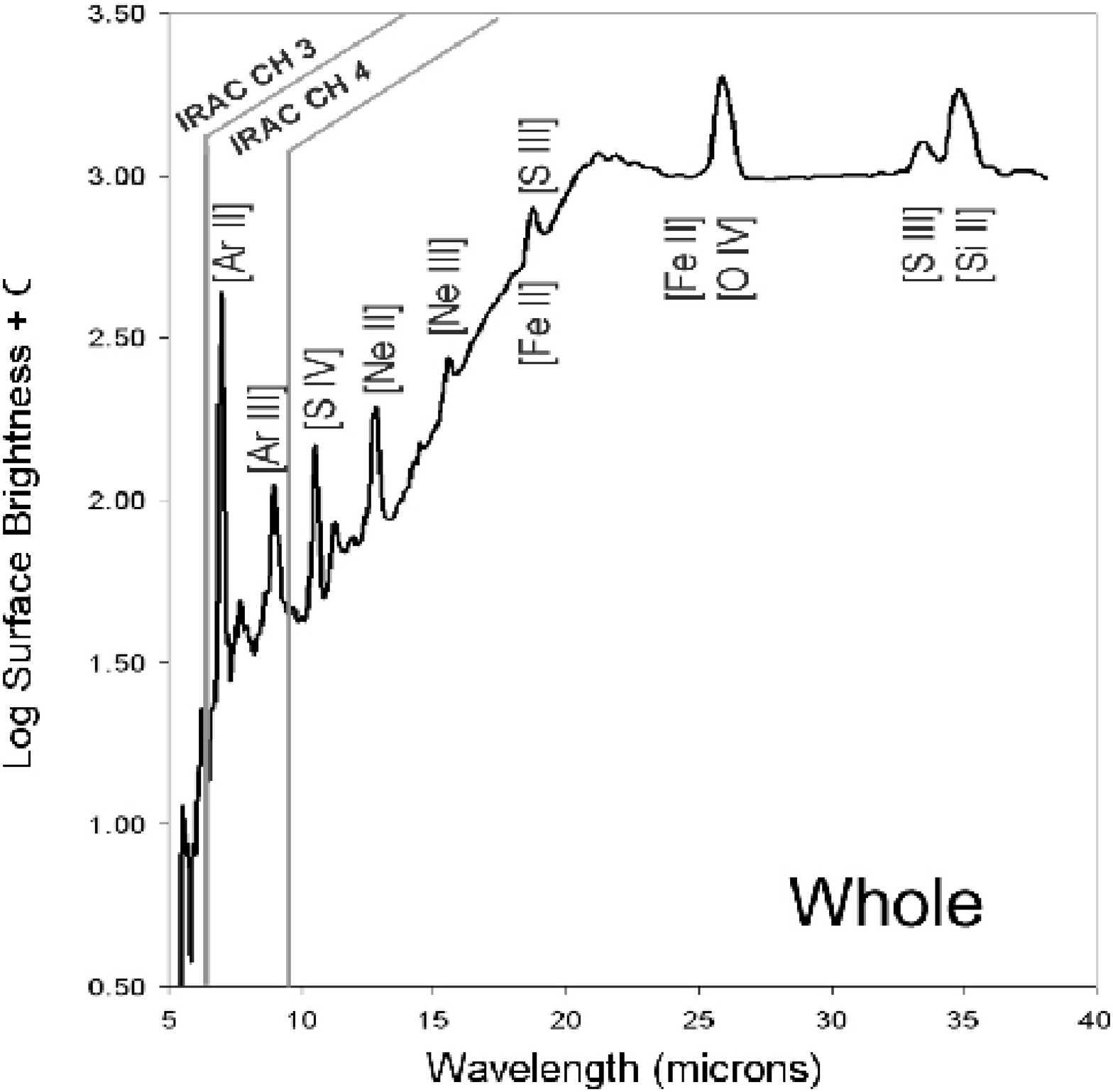}{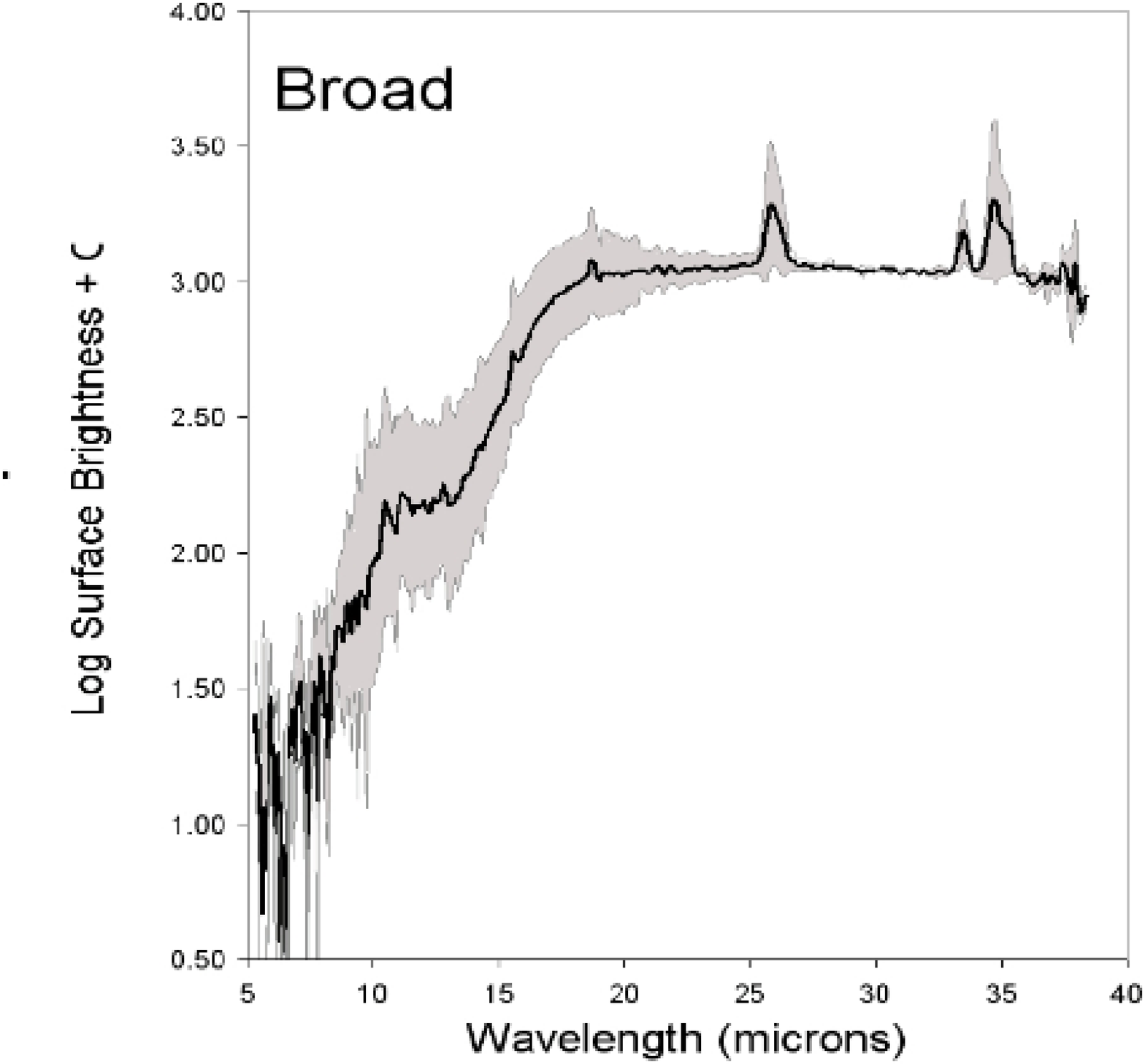}
\plottwo{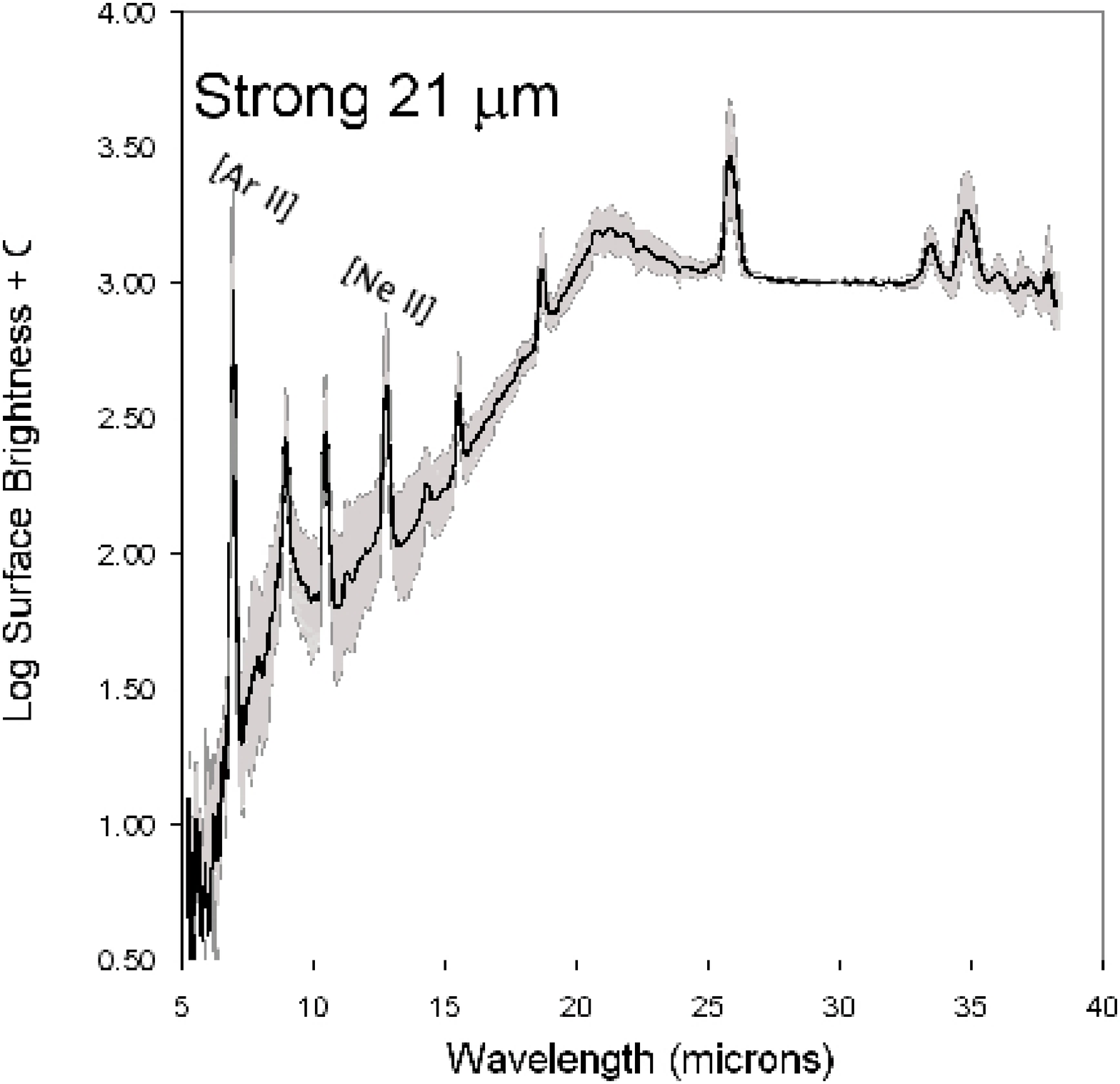}{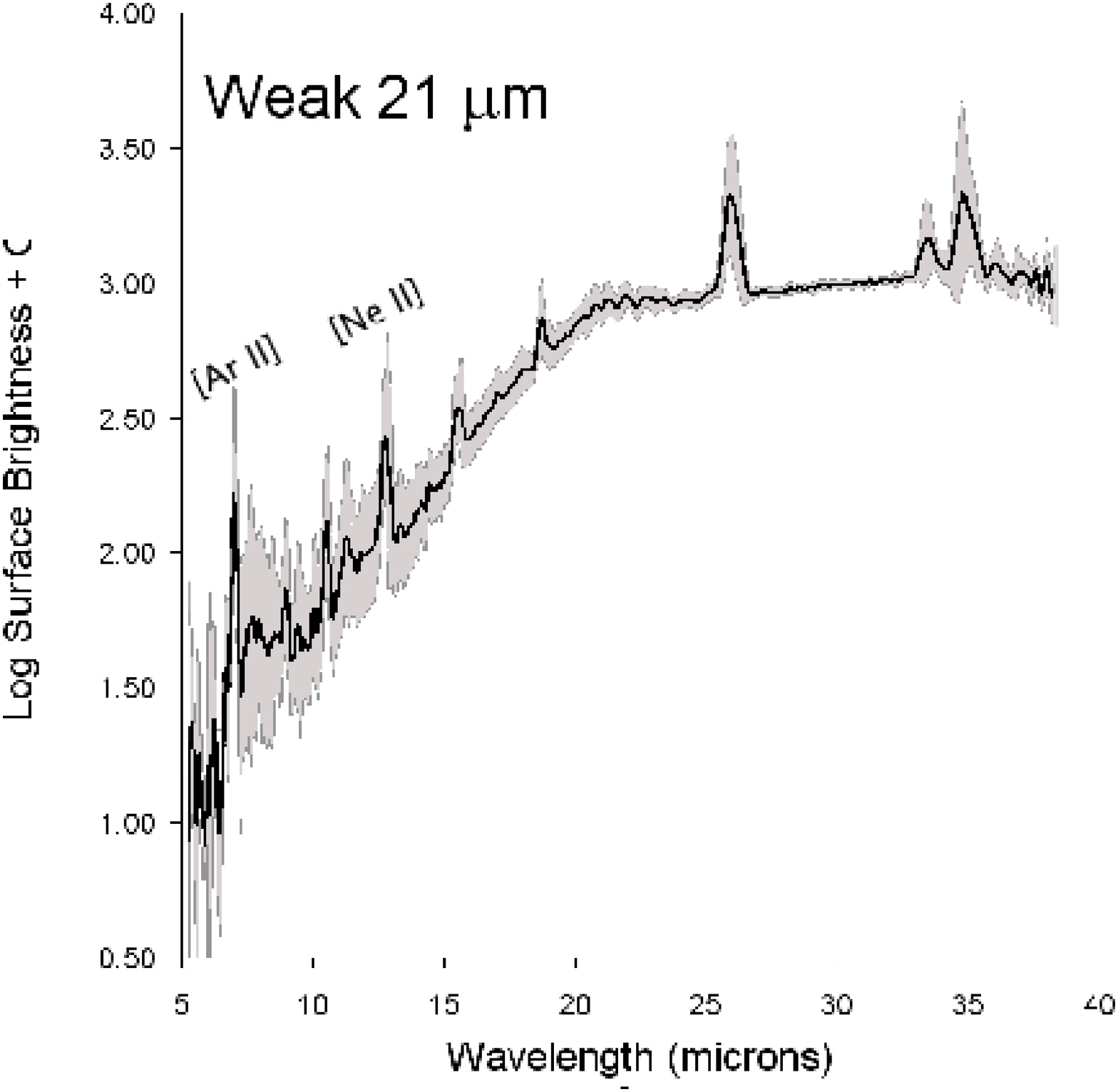}
\caption{IRS spectra of the total remnant and the three major spectral classes.
The spectra show the mean of the class, and the grey regions show the  rms scatter within the class. Within each class, all spectra are normalized to have the same brightness at 30~\micron. Small (up to $\approx$ 10\%) cosmetic adjustments have been made to normalize the SL and LL brightness scales.For full-resolution images, please see http://webusers.astro.umn.edu/\~{}jennis/iracpaper.html.}
\label{broadspec}
\end{center}
\end{figure}

\begin{figure}
\begin{center}
%\plotone{colorspec2.eps}
\plotone{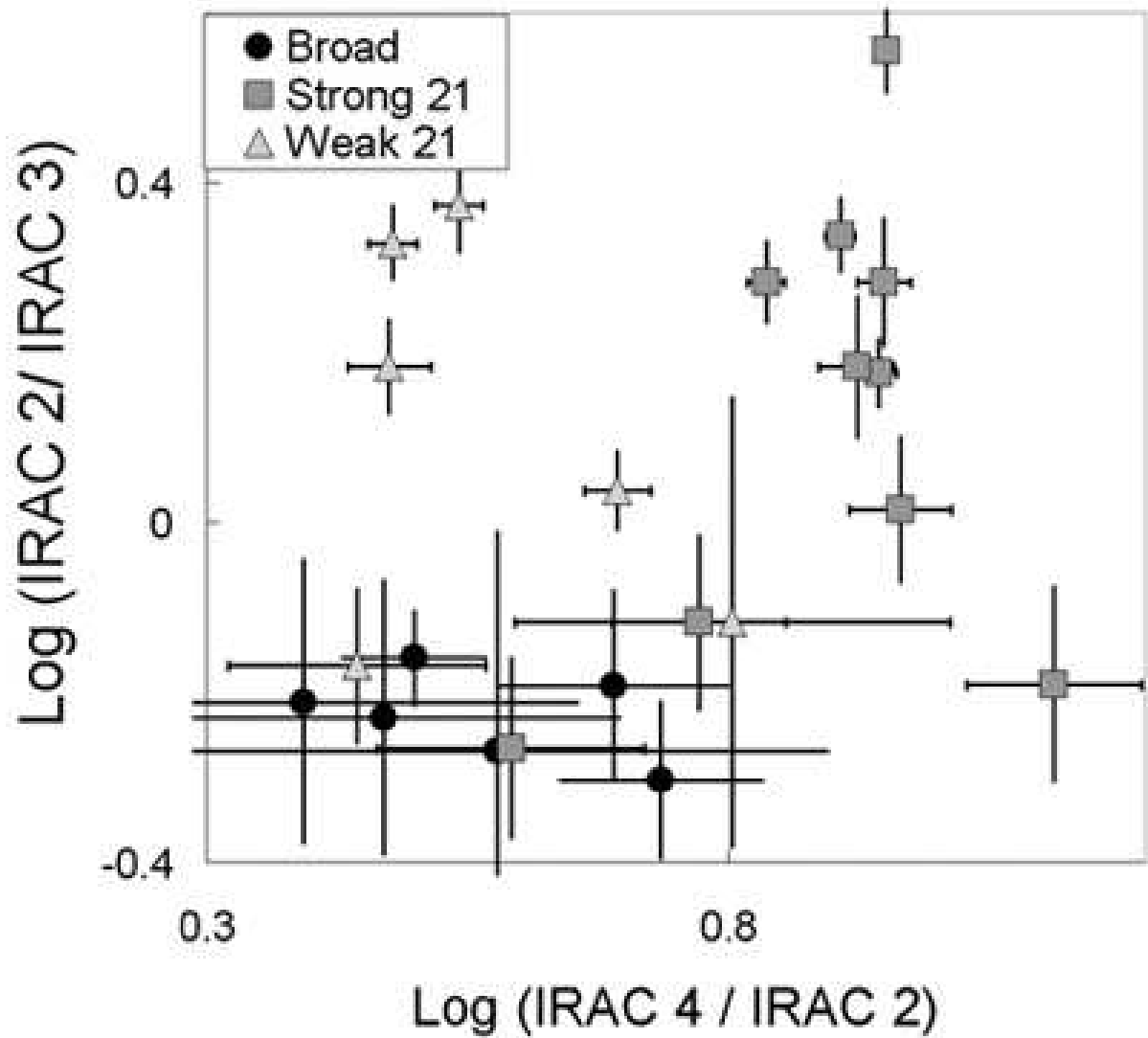}
\caption{IRAC colors for the three major classes of IRS spectra. For full-resolution images, please see http://webusers.astro.umn.edu/\~{}jennis/iracpaper.html.}
\label{colorspec}
\end{center}
\end{figure}

Channel 2, and to a lesser extent, Channel 3, have significant contributions from synchrotron radiation in some locations such as the northern forward shock region and interior filamentary structures (see Figure \ref{iracspec}).  We have therefore subtracted the Channel 1 image from Channels 2 and 3 to create the residual images in Figure \ref{channel23}.   The similarities between Channel 2 and Channel 4 are now more apparent, although the relative brightnesses of features vary by at least a factor of 5.  Channels 2, 3, and 4 all trace out the same large oval structure in the North, e.g., but their relative strengths vary abruptly between nearby knotty structures (see Figure \ref{iraccolor}). Sometimes, as in the far north of the bright ring, these changes are actually due to dynamically distinct but superposed features.

Bright Channel 2 regions show both strong 21~\micron~ and weak 21~\micron~ spectra, but when Channel 2 is brightest with respect to Channel 4, we find only weak 21~\micron~ spectra, along with low  [Ar II]/[Ne II] ratios (see Table 1).  These low ratios occur  because [Ar II] (Channel 4) is weak in these regions.   Most of the relatively bright Channel 3 regions  also have low [Ar~II]/[Ne II] ratios and weak 21~\micron~ spectra,  although some show modest peaks around 21~\micron.

There is no IRS coverage for Channel 2; we consider possible line contributions below.  Channel 3 is covered by the IRS from 5.3 to 6.4~\micron, but there is no coverage between 5.0 and 5.3~\micron. In the accessible  wavelength range we found no  ionic lines dominating the Channel 3 brightness, although there are very weak contributions in some locations from [Fe II] 5.3~\micron~ at the edge of the IRS band. To see where [Fe II] might be important for Channel 3,  we present the 18~\micron~ [Fe II] image in Figure \ref{channel23}.  [Fe II] emission is also seen at 1.64~\micron~\citep{rho} and in the spectra of some FMKs around 1.2~\micron~\citep{gerardyfesen}.  Comparison with our near infrared measurements suggests that [Fe II] might contribute up to 50\% of the Channel 3 emission in some isolated locations, but a negligible amount in many bright Channel 3 regions. The rest of Channel 3 emission is likely from the dust continuum.   When the synchrotron emission is subtracted from Channel 3, we also find a few isolated bright patches in the northeast jet and elsewhere (Figure \ref{channel23}). These bright patches are not coincident with features seen in Channels 2 and 4, and their IRS spectra  show the ``broad'' shape characteristic of the forward shock.

%In describing the associated broadband IRS spectral shapes  we use the terms ``weak-, intermediate-'' and ``strong-'' silicates, but  there is actually a continuous distribution in the strength of the 21~\micron~peak, relative, e.g., to the 26~\micron~continuum. This reflects in part the relative contributions of silicates and Al$_2$O$_3$ dust; some glassy carbon may also be required, especially in the ``weak-'' and ``intermediate-'' silicates spectra. These issues are discussed in more detail in  \cite{rho06}.
%The relative strengths of the [Ne II] and [Ar II] lines show a corresponding continuous distribution.

%ch3 (5-6.5um) [Fe II] (5.3um)-ISOSWS detected (JD plot), but weak.
%              dust (I think e.g. C dust): this explains why ch3 is strong in Ne moon.
%      So FeII at 53um will be very similar (80%) to that at near-IR.
%
%    1.64400  0.000365305 4D7 4F9      0.736119
%    5.34025  0.000289836 4F9 6D9      0.584043
%   a few position I measured show 50-70% may be Fe
%   But obviously, the rest 50-30% is somethingelse for the regions where
%Fe is present, but there are regions no Fe but strong ch3 (e.g. Ne moon),
%someother mechanism is needed. This is where I think the Caron dust present (unless
%some line related to Ne).

\subsubsection{What is Channel 2 (4 -- 5~\micron) ?}
In the absence of IRS coverage in the Channel 2 wavelength range, we summed the 17 available SWS spectra from the ISO archives (http://www.iso.esac.esa.int/ida) but found no strong lines shortward of the 6.99~\micron~[Ar II] line.  However, the fact that Channel 2 is usually brighter than both Channel 1 and Channel 3 (Figure \ref{iracspec}) indicates the presence of line emission.

There are two distinct questions regarding the origins of the Channel 2 emission -- what dominates when Channel 4  (and [Ar~II]) is also strong,  and what dominates when Channel 4 (and [Ar~II]) is weaker, but [Ne II] is still strong?  [Fe II] has several lines in the Channel 2, 4 - 5~\micron, band, but the 5.3~\micron~ emission is quite weak.  Looking at the 18~\micron~ [Fe II] structure, there is little or no emission where Channel 2 is strong relative to Channel 4 (see Figure \ref{moons} for the locations of these regions), so [Fe II] is unlikely to provide the missing lines in Channel 2.

If pieces of the hydrogen envelope survived the WR-wind stage of Cas~A, then Br~$\alpha$ at 4.05~\micron~could be present in the ejecta.  Since ground-based spectroscopy of Br~$\alpha$ is difficult, we obtained an image of the Pa~$\beta$ 1.28~\micron~ line using the Wide-Field Infrared Camera on the 5m Hale Telescope at Palomar Observatory. Some nearby [Fe II] lines, seen in FMKs by \cite{gerardyfesen} also fall into this filter. In a number of regions, the Pa~$\beta$, [Fe II] and Channel 2 images trace out the same structures, so Br~$\alpha$ might be responsible for some Channel 2 emission.  These are also the regions where Channel 4 is strong.  However, there are other locations where Channel 2 is strong, such as the jet and the crescent shaped regions seen in Figure \ref{moons}, that are weak or absent in the Pa~$\beta$ image, so Br~$\alpha$ is unlikely to be playing a key role.  We also see no evidence for Pf$\alpha$ at 7.46~\micron.

Another strong candidate is the CO fundamental bandhead around 4.76~\micron, such as seen in regions with shocked CO \citep{orionCO}. This bandhead has been detected in SN 1987A \citep{meikle,kotak}, as well as the first overtone at 2.29~\micron\ \citep[e.g.,][]{catchpole}, so we know that CO can form in supernova ejecta. 
% There are also a number of cold CO clouds exterior to the bright ring \citep{liszt}; if these were actually local to the remnant and interacting with Cas~A's
%shocks they could be responsible for the bright Channel 2 emission.
 The resolution of this issue requires sensitive spectra in the 2 - 5~\micron~band.

Another possibility is H$_2$, which we see in our IRS images around 17~\micron.  However, the H$_2$ is largely exterior to the remnant,  possibly associated with the surrounding CO clouds \citep{liszt}. If H$_2$ were dominant in Channel 2, it should also be strong both in 5-6.4~\micron~and 6.4-9.4~\micron~ spectra, but is not seen.

HeII 8-7 line recombination occurs at 4.76~\micron.  If this were responsible for Channel 2 emission, we should also have HeII 9-8 emission at 6.95~\micron~ and 10-9 emission at 9.71~\micron.  The former is unfortunately coincident with the extremely bright [Ar II] emission, and we find no evidence for the latter anywhere in the remnant.  At present therefore, the origin of the line emission in Channel 2 is unclear.

%[NOW PUT SOME MORE DETAILS ABOUT THIS:For a given spot, measure brightness in the Pbeta filter, erg/cm2/s/sr;
%scale by case B recombination to get Bralpha; convert units to MJy/sr; compare to the channel 2 surface brightness.
%Is it about right? The ratio of the two images could give the extinction, but for now let's just see if they are about right,
%and if the ratio channel2/Pbeta is equal to or a little higher than expected from CaseB and some extinction.
%I am game as to whether we make a longer discussion here featuring the Palomar image; Jeonghee can decide this part.]

%Some [\ion{Fe}{2}] lines are expected, but they should be much weaker than the 5.35, 17.93, and 25.99\,$\mu$m
%lines in the IRS spectra; we expect them to be a negligible contribution to IRAC Channel 2. Comparing the Channel 2 image
%to the locations of bright 17.9\,$\mu$m emission in the IRS spectra, we find no good correlation, so the evidence for an [\ion{Fe}{2}] contribution to Channel 2 is weak.

% Channel 4 continuum emission is anti-correlated in most places with the emission from Channel 3, as shown in Figure \ref{irac34}.  The flat-top spectra often rise around 23~\micron, so the weaker 21~\micron\ emission of flat-top spectra distinguishes them clearly from the 21~\micron\ peak spectra characteristic of Channel 4.  The strong Channel 4 regions show similar large scale structures to some seen in Channel 3, but in detail, there are often abrupt transitions to where one dominates over the other, such as on the southern bright ring or at the base of the jet in the NE (Figure \ref{irac34})

\section{Discussion}

We have found the mid-infrared radiation from Cas~A to arise from a number of different components - at short wavelengths,  synchrotron radiation and at longer wavelengths,  low ionization lines from Ne, O, Si, S, Ar and Fe ejecta, and shock-heated dust from both ejecta and CSM. The ejecta at different locations are further distinguished from each other by their colors in the IRAC bands, by the relative line strengths of different elements, and by the shape of their dust continua. The variations occur on both small and large spatial scales.  The same spatial  variations characterize the optical and X-ray line emission from ejecta, although these are from much higher ionization states.  In this discussion, we  briefly summarize the structure of  the multiwavelength appearance of the ejecta, and the  implications of these new Spitzer observations for the dynamics of Cas~A's explosion.

A consistent picture of the ejecta structure emerges from images at different wavelengths.  Line emission from elements such as Si and S dominate the optical \citep{fesenHST} and X-ray \citep{hwangholtpetre} emission, from moderate and high ionization states.  These appear structurally as a partially illuminated bright circular ring at the same location as the radio bright ring.  In addition, both optical and X-ray observations show the NE ``jet'' and  two interior elliptical rings towards the North. IRAC Channel 4 (6.4 - 9.4~\micron), which  has major contributions from [Ar II] and [Ar III] emission, shows all of these same structures.

\begin{figure}[t]
%\plotone{moonsbox.eps}
\plotone{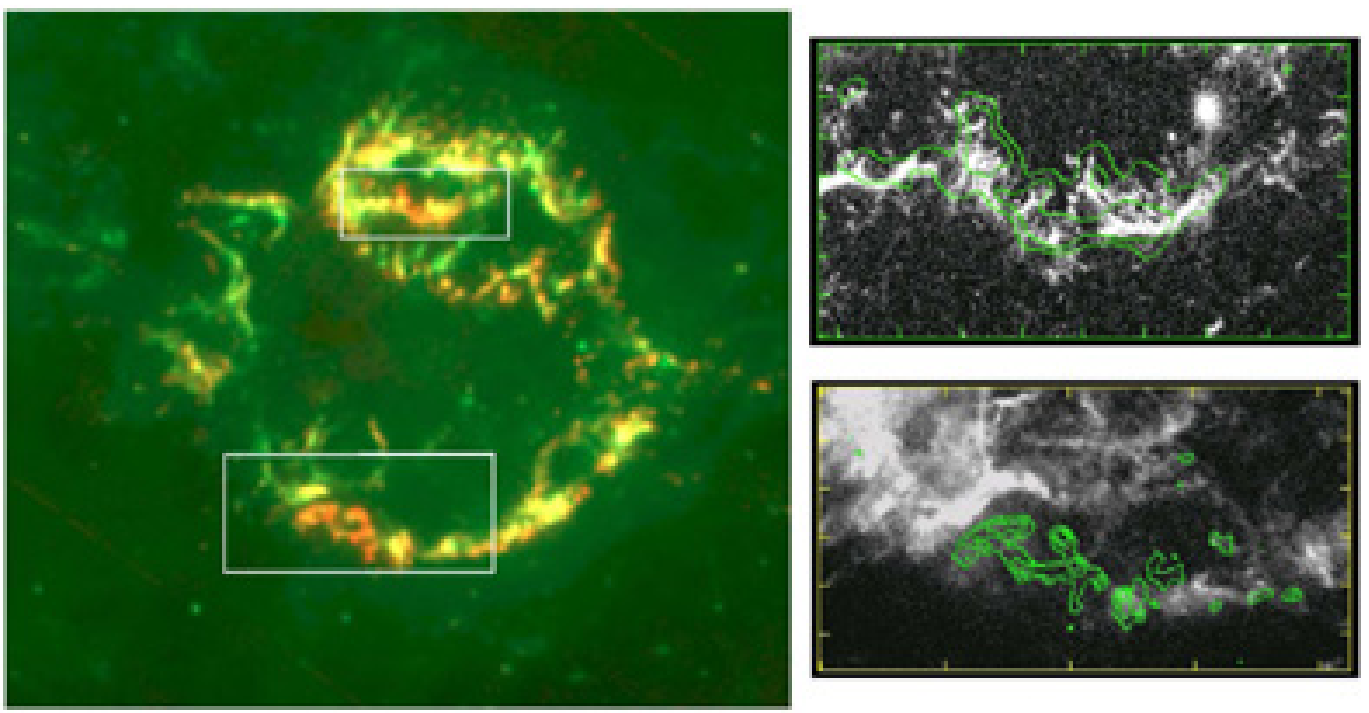}
\caption{Channel 2 (red) and Channel 4 (green).  The two red crescent regions (relatively high [Ne~II]/[Ar~II]) are shown on right, overlaid on images from other wavebands.  Top: Contours of Ch2/Ch4 ratio overlaid on the WFPC F450W image ([O~III]  $\lambda\lambda$4959, 5007) \cite{fesenHST}.  Bottom: Contours of Ch2/Ch4 ratio overlaid on Chandra Si image \citep{hwangholtpetre}. For full-resolution images, please see http://webusers.astro.umn.edu/\~{}jennis/iracpaper.html.}
\label{moons}
\end{figure}

A quite different picture of the ejecta emerges from IRAC  Channel 3 (5 - 6.4~\micron), which is dominated by dust, and in regions where Channel 2 (4 - 5~\micron) is strongest with respect to Channel 4.  Regions of high (Channel 2 / Channel 4) can be seen in Figure \ref{moons}, where the most prominent features are  two bright crescent features.   The ratio of [Ar II]/[Ne II] from our IRS spectra (see Table 1) is low in these regions, so they are relatively Ne-rich, although Ne does not directly contribute in the IRAC bands.  The northern crescent is also shown in Figure \ref{moons} overlaid on the WFPC2 F450W image, showing that the same crescent structure appears in  [O~III] $\lambda\lambda$ 4959, 5007 emission.  This feature is also seen in an X-ray image in the oxygen emission around  0.64-0.71~keV, using unpublished ACIS data from our proper motion studies \citep{delaney04}. The brightness of the northern crescent in the   F850LP WFPC2 image, which is primarily sensitive to  [S III] $\lambda\lambda$ 9069, 9531 emission, is a factor of two lower, compared to the [O~III] emission,  than in surrounding regions. The southern crescent falls in a very distinct X-ray gap, as seen in Figure \ref{moons}.  Multiple epoch HST images show brightening [O III] emission at this location (R. Fesen, private communication). Although there is some X-ray  Si, S, or Fe emission in the general area of the crescents \citep{hwangholtpetre}, they do not show the detailed structural correspondence seen in the O X-rays.

The ejecta thus appear to spatially segregate in a way consistent with  different nucleosynthetic layers.  The brightest optical, X-ray and 8~\micron~ (Channel 4) emission is dominated by lines from the O-burning layers, e.g.,  Si, S and Ar.   Other regions, such as the crescents, are distinguished  by their relatively  strong 4 - 6~\micron~ emission and the relatively strong products of C-burning, neon (infrared) and oxygen (optical and X-ray).  This suggests that the appearance of different elements at different locations in the remnant reflects which nucleosynthetic layer is locally illuminated.

This picture receives strong support from the different types of dust found associated with regions of different IRAC colors.
Where Channel 4 is strong, the IRS spectra show  a broad spectrum with a  21~\micron~peak, which requires the presence of silicates, e.g. MgSiO$_3$ or Fe or Mg proto-silicates \citep{lagage, arendtiso, rho06}. Thus, the O-burning product Si is also seen in the dust.   Where Channels 2 and 3 are strongest relative to Channel 4, the gently rising weak 21~\micron~ spectra are dominated  by Al$_2$O$_3$, i.e.,  the C-burning products.  The detailed models of dust temperature and composition, including contributions from carbon dust, are presented in \cite{rho06}.

%This picture is further reinforced by comparing the regions of relatively strong Channels 2,3 to optical and X-ray images.  Figure \ref{moons} shows the two brightest, crescent-shaped  regions in the (Channel 2/Channel 4) ratio image.  In the South, the Channel 2,3 crescent fills the distinct gap seen in all X-ray line images \citep{hwang, vlp}. Little or no emission is seen at this location in the HST WFPC2 images \citep{fesenhst}.  Thus, in this location, we find little or no evidence for shock-heated products of Si, S, Fe, Ar in the infrared, optical or  X-ray, but do find emission from [Ne II] and  Al$_2$O$_3$ dust.   Thus, we have optical, X-ray, and infrared evidence for the presence of Ne, O, and Al, all C-burning products, and no evidence for O-burning products at this location.

These findings provide a new perspective on the distribution of different elements on Cas~A's sparsely-covered spherical shell, which appears in projection as the bright ring.  Apparent differences in composition at different locations could have appeared from variations in temperature and/or ionization states.  Similarly, variations in ionization timescale \citep{hwanglaming} and density can significantly affect which elements are seen.   However, the results presented here indicate that differences in apparent composition likely reflect the actual local composition.  In some locations, we find multiple indicators for only  C-burning products, {\it e.g.} neon, and oxygen in the form of [O III], [O VIII] and Al$_2$O$_3$ dust. In other locations we see  O-burning products.  For example, when sulfur is seen, e.g., it appears as  [S III] (33~\micron, see Fig. \ref{broadspec}) and He- and H- like S  (X-ray S~XV and S~XVI lines between $\approx$2.4 and 3.1 keV).  Silicon, where it is present, appears as strong 21~\micron~ dust (i.e., silicates),   [Si II] (35~\micron), and He-  and H- like Si (X-ray Si~XIII and Si~XIV lines between $\approx$1.8 and 2.6~keV).
These variations in composition at different locations likely reflect asymmetries in the original explosion.
%The Ne-rich crescents, for example,  are symmetrically positioned around the X-ray point source \citep{Smith06} within a few arcseconds, and roughly perpendicular to the strong jet/counterjet axis \citep{hwangmsec}.

% The inferred proper motion of the point source is along the same direction, (169$^o\pm$8.4$^o$) \citep{fesen06} as the line connecting the crescents (167$^o\pm$3$^o$)In these same directions there are distinct gaps in the distribution of 1825 high velocity  N, O, and S outlying knots recently measured with HST \citep{fesen06}.  It thus seems very likely that we are seeing a bipolar dynamical structure linked to nucleosynthetic layers in the explosion. A further dynamical analysis is presented in \cite{Smith06}.

\begin{figure}
\begin{center}
%\plotone{caslayers4.eps}
\plotone{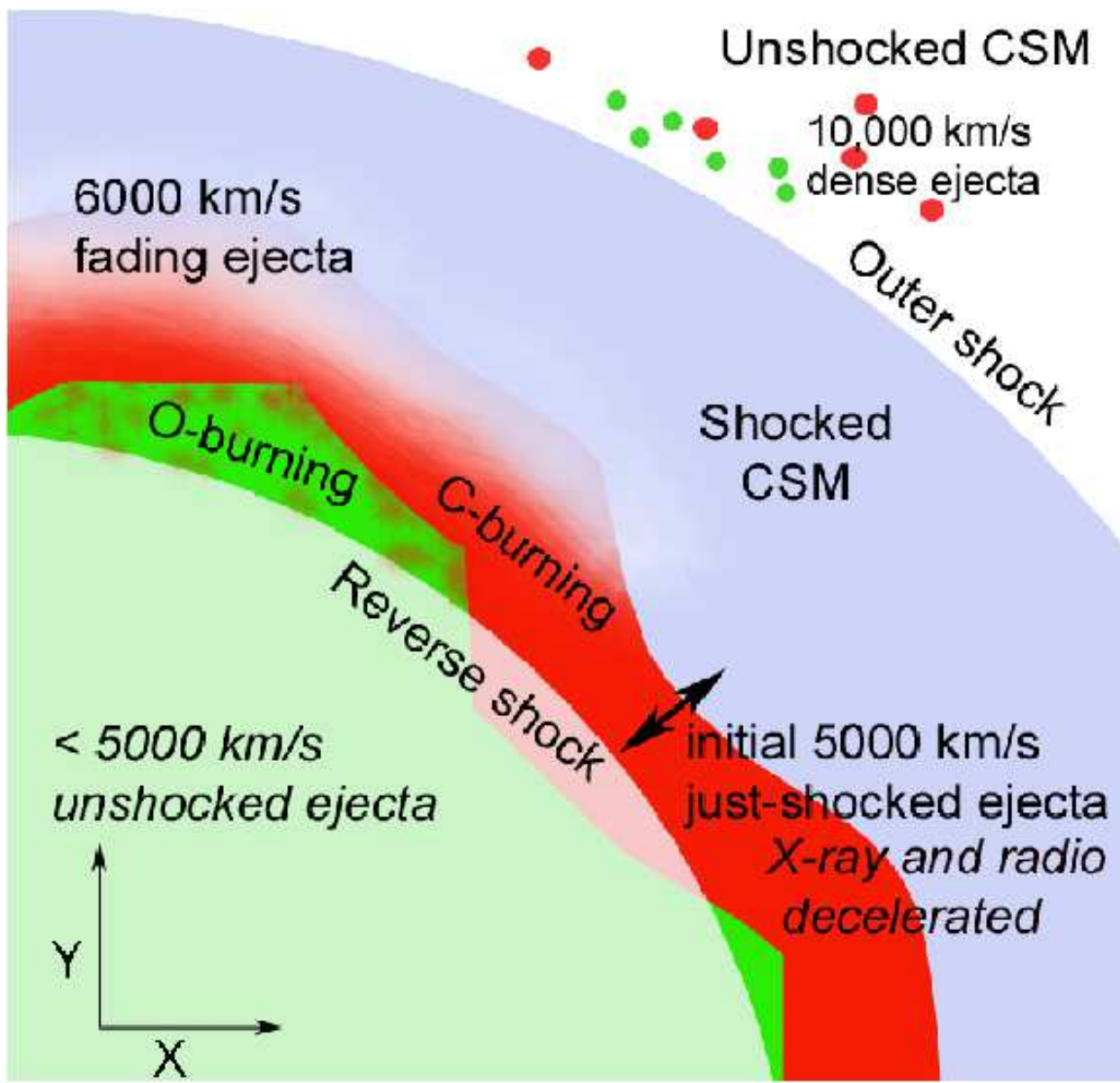}
\caption{Illustration of dynamical model described in text. In different directions, different layers from the explosion are currently reaching the reverse shock. There, they develop internal shocks and become visible for a short time across all wavebands before fading. X-ray and radio emission do not appear until significant deceleration has occurred. For full-resolution images, please see http://webusers.astro.umn.edu/\~{}jennis/iracpaper.html.}
\label{illustrate}
\end{center}
\end{figure}

 We briefly outline a simple dynamical model for Cas~A which incorporates these findings.   The blast wave from the explosion was quite symmetric, as seen by the nearly circular appearance of the outer and global reverse shocks \citep{gotthelf}, and has now  swept up sufficient mass to decelerate the outer shock by a factor of $\approx$1.5 \citep{delaney04} from free expansion.  This produces an inward-moving (in the frame of the explosion) global reverse shock in the diffuse X-ray gas. Clumps of ejecta traveling at $\approx$5000~km/s will be encountering this reverse shock at the current epoch, driving a  {\it local} reverse shock back into the clumps. This heats and compresses  them,  making them visible optically \citep{morse}. Stripping, heating, ionization and disruption of these clumps then lead to decelerated X-ray and radio-emitting features \citep{delaney04,anderson94} and eventual disappearance. The reverse shock then is a slowly moving front that is successively overtaken and illuminated as a bright ring.  by increasingly slower-moving undecelerated ejecta from the initial explosion.  An illustration of the key features of this picture is shown in Figure \ref{illustrate}.

% Dense clumps of ejecta were produced in the explosion over a broad range of velocities, $<$5000 - 15000~km/s.  As these clumps encounter a density discontinuity, a , in  patchy, often ring-like structures around the remnant.  While still radiative, they experience only small decelerations from free expansion \citep{fesen06}.  This describes the  overall symmetric aspects  of the explosion, with a new generation of slower moving dense ejecta clumps continuously encountering the global reverse shock while still in free expansion, and keeping the bright ring illuminated with new material.

Superposed on this symmetric structure are major variations, the most prominent being the jet and counterjet regions.  They arise deep in the explosion, producing fast-moving S-rich ejecta \citep{fesen06}, as well as emission from Si group elements \citep{hwangmsec}.   Roughly perpendicular to this axis, we now find evidence for a much slower moving bipolar structure, in the form of the crescents of C-burning material in the infrared, optical, and X-ray.  In order to reach the global reverse shock and become visible now, this material must have moved at a free expansion speed of $\approx$5000 km/s.  However, along this axis only these upper layers are now encountering the reverse shock. Material from the O-burning layers is not seen either at the reverse shock, or in the fast-moving outlying knots in these directions \citep{fesenN}.

This scenario leads to the suggestion that if we could wait sufficiently long, we would see the O-burning layers encounter the reverse shock along the crescents' axis, and become visible at all wavelengths. We find evidence that this actually may be occurring, because of the presence of [Si II] and [S III] emission interior to and somewhat overlapping the bright ring.

\section{Conclusions}

1. The four IRAC bands dominate in different regions of the Cas~A supernova remnant, echoing structures seen in optical, X-ray and radio images.

2. IRAC Channel 1 is dominated by infrared synchrotron radiation;  where Channel 1 dominates, the broadband spectra have a distinct shape, gently peaking around 26~\micron, which we attribute to forward-shock-heated circumstellar dust.

3. IRAC Channel 4 has a significant contribution from  both [Ar II] emission and continuum.  Where Channel 4 dominates, the dust continuum peaks strongly around 21~\micron, signifying the presence of silicates.

4. Where IRAC Channels 2 and 3 are strongest with respect to Channel 4, [Ar II] is weaker relative to [Ne II]. The continuum in these regions rises slowly or levels off at 21~\micron, showing the absence or only weak presence of silicate dust.

5. The relatively strong Channel 2,3 regions  show  optical and X-ray oxygen emission and an absence of silicon and sulfur.

6. We present a dynamical picture for the Cas~A's explosion in which cones of ejecta are produced with different velocities in different directions, In some directions, only the upper C-burning layers have reached the reverse shock, while in other directions, deeper O- and Si- burning layers have done so.  A particularly interesting bipolar structure, perpendicular to the jet/counterjet axis, has been identified.

\acknowledgements{This work is based on observations made with the Spitzer Space Telescope, which is operated by the Jet Propulsion Laboratory, California Institute of Technology under a contract with NASA. Support for this work was provided by NASA through an award issued by JPL/Caltech.}

\clearpage

\clearpage

%begin page of six images
%13

%\begin{figure}
%\begin{center}
%\plotone{Totalspec3.eps}
%\plotone{f2.eps}
%\caption{Low spectral resolution IRS spectrum of Cas~A, covering the entire bright ring and portions of the outer shock and jet,  The ripples at the 21~\micron~ peak are instrumental in origin.  All lines have been heavily Doppler broadened.  The lines around 18.5~\micron~ and 26~\micron~ are blended here. The wavelength regions covered by IRAC Channels 3 and 4 are indicated. There is no corresponding IRS coverage for IRAC Channels 1 and 2.}
%\end{center}
%\label{total}
%\end{figure}

%\begin{figure}[t]
%\begin{center}
%\includegraphics[width=8cm]  {ringspeclab04.eps}
%\includegraphics[width=8cm] {csmspeclab.eps}
%\includegraphics[width=8cm] {nemoonspeclab04.eps}
%\includegraphics[width=8cm]{totalspeclab04.eps}
%\includegraphics[width=8cm] {f5a.eps}
%\includegraphics[width=8cm] {f5b.eps}
%\includegraphics[width=8cm] {f5c.eps}
%\includegraphics[width=8cm] {f5d.eps}
%\end{center}
%\caption{IRS spectra from the three regions indicated in Figure \ref{iraccolor}, representing the major classes of observed spectra, plus the ``total'' spectrum, which includes the bright ring, its interior, and portions of the forward shock and jet, The brightness scale is logarithmic. Small cosmetic adjustments have been made to normalize the SL and LL brightness scales.  }
%\label{broadspec}
%\end{figure}

%\begin{figure}[t]
%\begin{center}
%\includegraphics[width=9cm]{RG14.eps}
%\includegraphics[width=9cm]{f7.eps}
%\end{center}
%\caption{Color combination showing IRAC Channel 1 (red) and  Channel  4 (green). In most locations, the colors effectively separate synchrotron  emission from behind the forward shock (red) from green/yellow regions of reverse-shocked ejecta.}
%\label{irac14}
%\end{figure}

\clearpage
\begin{table}
\caption{Characteristics of emission for locations where each IRAC channel is enhanced relative to the other channels.}
\vskip 0.25in
\begin{tabular}{|c|c|c|c|c|c|}
\hline
Channel &   Emission           &  [Ar II]/[Ne II]   & Broadband shape      &       Origin \\
 $\lambda\lambda$ \micron~ &   &           &        (Fig. 2 color)   &     \\  \hline
 1          &    {\bf  synchrotron}  &     no lines            &  Broad          &  Forward shocked    \\
3.2 - 3.9      &               &            &  (blue)     & CSM          \\ \hline
 2         &      CO?, Br~$\alpha$?          &    1.8         &  Weak 21~\micron      &   C-burning products   \\
4.0 - 5.0    &   [Fe II]?, dust           &       &    (green)   &   (O, Si- burning \\
&          &                           &  &~ ~ ~ products) \\ \hline
 3      &        {\bf dust}, [Fe II]        &    1.7         & Weak 21~\micron    &  C-burning products       \\
  5.0 - 6.4  &         &               &     (orange/yellow)    & (O, Si- burning  \\
&          &                           &   & ~ ~ ~ products)\\ \hline
 4      &   {\bf [Ar II]}, [Ar III]       &    3.7       &  Strong 21~\micron              &  O, Si burning\\
 6.5 - 9.4  & dust          &                 &   (red)       &  products          \\\hline
%synchrotron & forward-shock-heated CSM & broad 26~\micron\ & Very few, some [OIV],[FeII] at 26~\micron\ & not newly formed\\ \hline
%3 & continuum & continuum &
%4 &
%2 & [FeII], P$\beta$, synchrotron & Where stronger than Ch 4, Stronger [NeII], [ArII] & Flat-top and %21~\micron\ peak & Where stronger than Ch 4, Stronger [NeII], [ArII] & Ne?\\ \hline

\end{tabular}
\vskip .15in \noindent Notes:  The primary contributor to each IRAC channel is shown in {\bf boldface}.
\end{table}

\vskip 0.25in

\end{document}